\documentclass[10pt,conference]{IEEEtran}

\IEEEoverridecommandlockouts
\usepackage{cite}
\usepackage{amsmath,amssymb,amsfonts}
\usepackage{algorithmic}
\usepackage{graphicx}
\usepackage{textcomp}
\usepackage{xcolor}

\usepackage[ruled,vlined,linesnumbered]{algorithm2e}
\usepackage{multirow}
\usepackage{adjustbox}
\usepackage{pifont}
\usepackage{threeparttable}
\usepackage{booktabs}
\usepackage{bbding}
\usepackage{tablefootnote}
\usepackage{hyperref}
\usepackage{tcolorbox}
\usepackage{enumitem}
\usepackage{enumerate}
\usepackage{subfig}
\usepackage{url}
\usepackage{soul}
\usepackage{balance}
\usepackage{makecell}
\hypersetup{
colorlinks=true,
urlcolor=black,
linkcolor=black,
citecolor=black
}

\usepackage{listings}

\def\BibTeX{{\rm B\kern-.05em{\sc i\kern-.025em b}\kern-.08em
    T\kern-.1667em\lower.7ex\hbox{E}\kern-.125emX}}

\begin{document}

\newcommand{\found}{12}

\newcommand{\tool}{\textsc{r}TED}
\newcommand{\tech}{\textsc{r}TED}

\newcommand{\noanalysis}{\tool$_{w/o\text{ }c}$}
\newcommand{\noreflection}{\tool$_{w/o\text{ }r}$}

\newcommand{\erroranalysis}{error-seeking agent}
\newcommand{\normalanalysis}{non-error-seeking agent}

\newcommand{\pyinder}{Pyinder}
\newcommand{\tester}{CHATTESTER}
\newcommand{\sym}{SymPrompt}
\newcommand{\hits}{HITS}

\newcommand{\homepage}{https://anonymous.4open.science/r/TODO}

\newcommand{\xmark}{\ding{55}}
\newcommand{\cmark}{\ding{51}}

\newcommand{\jj}[1]{{\color{orange}[Junjie: #1]}} 
\newcommand\yc[1]{\textcolor{red!50}{#1}}
\newcommand{\jiang}[1]{\textcolor{blue}{#1}}

\newcommand{\colorblue}{\color{black}} %
\newcommand{\ins}[1]{{\colorblue{#1}}}

\newcommand{\del}[1]{{\color{red}{#1}}}

\author{
 \IEEEauthorblockN{
    Chen Yang\IEEEauthorrefmark{4}, 
    Ziqi Wang\IEEEauthorrefmark{4}, 
    Yanjie Jiang\IEEEauthorrefmark{4}, 
    Lin Yang\IEEEauthorrefmark{4}, 
    Yuteng Zheng\IEEEauthorrefmark{4}, 
    Jianyi Zhou\IEEEauthorrefmark{2}, 
    Junjie Chen\IEEEauthorrefmark{4}\IEEEauthorrefmark{1}
     \thanks{\IEEEauthorrefmark{1} corresponding author}
}
\IEEEauthorblockA{
    \IEEEauthorrefmark{4}College of Intelligence and Computing, Tianjin University, Tianjin, China\\
    \IEEEauthorrefmark{2}Huawei Cloud Computing Technologies Co., Ltd., Beijing, China
}
\IEEEauthorblockA{
    \{yangchenyc, wangziqi123, yanjiejiang, linyang, zyt\_767904, junjiechen\}@tju.edu.cn, zhoujianyi2@huawei.com
}
}

\title{Reflective Unit Test Generation for Precise Type Error Detection with Large Language Models} 

\maketitle

\begin{abstract}

Type errors in Python often lead to runtime failures, posing significant challenges to software reliability and developer productivity. Existing static analysis tools aim to detect such errors without execution but frequently suffer from high false positive rates. Recently, unit test generation techniques offer great promise in achieving high test coverage, but they often struggle to produce bug-revealing tests without tailored guidance.
To address these limitations, we present \tool{}, a novel type-aware test generation technique for automatically detecting Python type errors. Specifically, \tool{} combines step-by-step type constraint analysis with reflective validation to guide the test generation process and effectively suppress false positives.
We evaluated \tool{} on two widely-used benchmarks, BugsInPy and TypeBugs. 
Experimental results show that \tool{} can detect 22$\sim$29 more benchmarked type errors than four state-of-the-art techniques.
\tool{} is also capable of producing fewer false positives, achieving an improvement of 173.9\%$\sim$245.9\% in precision.
Furthermore, we applied \tool{} to six real-world open-source Python projects, and successfully discovered \found{} previously unknown type errors, demonstrating \tool{}’s practical value.

\end{abstract}

\begin{IEEEkeywords}
Test Generation, Type Error, Bug Detection
\end{IEEEkeywords}

\section{Introduction}
\label{sec:intro}

Type errors are among the most common and critical issues in Python applications. They can lead to unexpected crashes and pose significant risks to the reliability of systems across various domains, including artificial intelligence platforms, data science pipelines, and financial applications.
Despite their severity, type errors remain highly prevalent. According to Oh et al.~\cite{pyinder}, they account for over 30\% of Python-related questions on Stack Overflow and GitHub issues, underscoring the urgent need for effective techniques to detect such errors.

To address this, several static techniques have been proposed for detecting type errors~\cite{pyinder, pyright}. However, these techniques often suffer from high false positive rates, which significantly limit their practicality. 
For example, PyInder~\cite{pyinder}, one of the most advanced static type checkers, reported tens of thousands of warnings but identified only 34 real type errors. This overwhelming noise renders such tools impractical for developers, who cannot afford to spend excessive time triaging false alarms.
In contrast, unit testing is generally considered to produce fewer false positives in software quality assurance~\cite{auger}.
However, existing unit test generation techniques (including search-based~\cite{pynguin} and LLM-based techniques~\cite{chattester, symprompt, hits}) typically use code coverage as their testing guidance, which is not directly aligned with the goal of bug detection and may thus limit their effectiveness in uncovering bugs (particularly for certain types of bugs such as type errors).
Indeed, prior studies have shown that high coverage does not necessarily translate into effective bug detection~\cite{npe, pynguin, chen2019compiler}.
Also, as demonstrated in our empirical study (presented in Section~\ref{sec:rq1}), the state-of-the-art unit test generation technique (i.e., \hits{}~\cite{hits}) was only able to detect 12 type errors out of the 69 benchmarked bugs.

Intuitively, shifting the testing guidance from enhancing code coverage to directly targeting bug detection could be helpful.
Particularly, this shift is more feasible for LLM-based unit test generation, which can be guided through lightweight prompting, compared to search-based techniques that typically require significant engineering effort.
Moreover, search-based techniques could struggle with generalizability across Python’s diverse versions and its evolving type system.
Therefore, our work focuses on leveraging LLM-based unit test generation to enhance the detection of type errors.
However, simply instructing LLMs to generate unit tests for detecting type errors does not yield satisfactory effectiveness.

\begin{figure*}[t]
  \centering
  \includegraphics[width=1\linewidth]{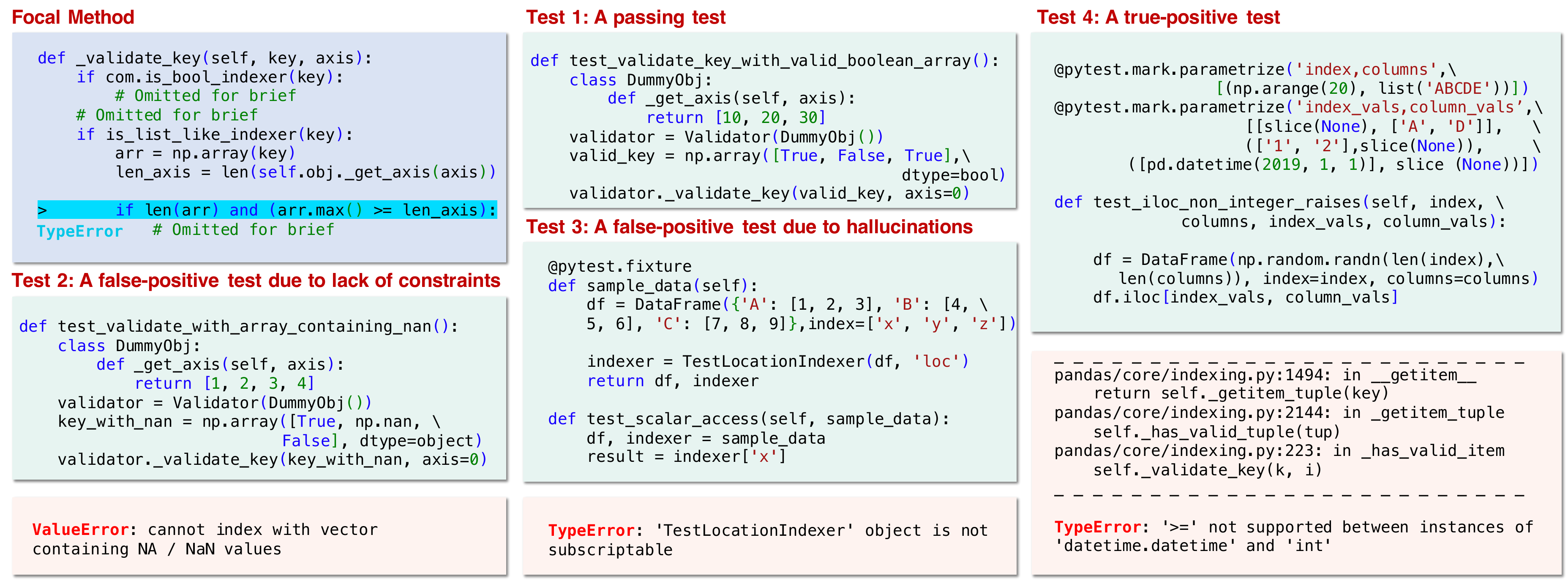}
  \caption{Motivating example}
  \label{fig:motivating_example}
\end{figure*}

On one hand, LLMs are primarily trained on non-buggy code paired with passing tests, which leads them to favor safe and common usage patterns and test inputs, thereby limiting their ability to uncover type errors.
While explicitly prompting the LLM to generate bug-revealing inputs may seem like a viable solution, on the other hand, it suffers from another challenge of context-aware type constraints in Python.
Specifically, LLMs often lack awareness of such constraints, which typically arise from broader usage contexts that are not evident when analyzing the focal method in isolation.
Violating these constraints leads to invalid crashes that do not reveal type errors, but instead indicate improper or unsupported method invocations.
For example, as shown in Fig~\ref{fig:motivating_example} (Test 2), the LLM generates a unit test for the method \texttt{\_validate\_key}.
This test passes a {\tt NumPy} array containing a {\tt NaN} value to the method, resulting in a crash. 
However, in actual usage, the input is first validated by the method \texttt{has\_valid\_item}, which ensures that each element supports certain required magic methods, such as \texttt{\_\_bool\_\_}.
Since \texttt{NaN} lacks these methods, it violates the context-aware constraint that the LLM fails to capture when focusing solely on the focal method.
This leads to an invalid crash — effectively a false positive in type error detection.
Moreover, LLM hallucinations can further exacerbate the issue of false positives.

To achieve precise type error detection, we propose \textbf{\tool{}} (\textbf{R}eflective \textbf{T}ype \textbf{E}rror \textbf{D}etection with LLMs), a novel technique that guides LLM-based unit test generation with type constraint analysis and self-reflection.
First, \tool{} captures context-aware type constraints by analyzing the invocation chains leading to the focal method, enabling step-by-step backward propagation of these constraints.
To enhance its effectiveness in uncovering type errors, \tool{} incorporates an \erroranalysis{} that performs type constraint propagation analysis: 
it identifies input types likely to trigger type errors in the focal method and infers the corresponding constraints that earlier methods in the invocation chain must satisfy to produce these inputs.
By taking the inferred constraints, the invocation chain, along with the context information of the entry method as prompt, \tech{} then guides the LLM to generate bug-revealing tests for the invocation chain.

Due to the hallucination issues inherent in LLMs, the triggered failures may correspond to either true bugs or false positives. 
To mitigate the impact of false positives, \tech{} incorporates a self-reflection mechanism comprising three specialized agents: a type-consistency-checking agent, a semantic-validity-checking agent, and a meta-evaluation agent.
The type-consistency-checking agent verifies whether the input types of the generated test conform to the constraints derived from type constraint analysis.
The semantic-validity-checking agent assesses whether the generated test adheres to semantic expectations, such as contextual constraints or usage patterns that would not occur in realistic scenarios.
The meta-evaluation agent integrates the judgments from the other two agents to determine whether a failure represents a genuine type error or a false positive. 
If a false positive is inferred, \tool{} uses this reflective feedback to refine its test generation process, enabling iterative improvement and enhancing the precision of the generated tests.

We evaluated \tool{} on two widely-used benchmarks in this field, i.e., BugsInPy~\cite{bugsinpy} and TypeBugs~\cite{pyter}, involving 16 Python projects with 69 real-world type errors.
We compared \tool{} with
the state-of-the-art Python type error detection technique (i.e., \pyinder{}~\cite{pyinder}) and the state-of-the-art LLM-based unit test generation techniques (i.e., \tester{}~\cite{chattester}, \sym{}~\cite{symprompt}, and \hits{}~\cite{hits}). 
Specifically, \tool{} detects 34 type errors with 13 false positives, while \pyinder{}, \tester{}, \sym{}, and \hits{} detect only 5, 7, 3, and 12 bugs, with 21, 19, 14, 35 false positives, respectively. 
That is , \tool{} is not only capable of detecting 22$\sim$29 more type errors than compared techniques, but also achieves an improvement of 173.9\%$\sim$245.9\% in precision.
Furthermore, we applied \tool{} to the latest versions of six popular open-source Python projects, and discovered \found{} previously unknown type errors. 
These results demonstrate the effectiveness of \tool{} in uncovering type-related bugs through generating effective unit tests.

In summary, our contributions are as follows:
\begin{itemize}
    \item 
    We design \tool{}, a novel technique that leverages context-aware type constraint analysis combined with a self-reflection mechanism to guide LLM-based unit test generation, enabling precise Python type error detection.
    
    \item
    \tool{} introduces a type constraint analysis method that captures context-aware constraints via invocation chain analysis and employs an error-seeking agent to propagate constraints backward step-by-step, guiding the generation of targeted test inputs more likely to expose type errors.
    
    \item 
    \tool{} incorporates a self-reflection process consisting of three specialized agents — type consistency checking, semantic validity checking, and meta-evaluation — to validate generated failures and iteratively refine test generation, reducing false positives in type error detection.

    \item We evaluated \tool{} on real-world Python projects, significantly outperforming state-of-the-art baselines on benchmarked bugs and detecting \found{} previously unknown type errors in the wild.
\end{itemize}

\section{Motivation}
We illustrate the challenges of detecting Python type errors with a motivating example and describe how our technique mitigates them. The motivating example is presented in Figure~\ref{fig:motivating_example}, sampled from the real-world open-source project Pandas~\cite{pandas}. In this example, the type error is triggered by the operation {\tt arr.max() $\ge$ len\_axis}, where incompatible operand types can lead to a runtime crash
({\tt len\_axis} is an {\tt int}, but {\tt arr.max()} may return a non-integer type depending on the {\tt key} argument).

To detect this bug, we need a test that exercises the faulty code. First, we used \tester{}'s default prompt to instruct the state-of-the-art LLM, DeepSeek-V3~\cite{deepseek}, to generate a test, which we refer to as \texttt{Test 1}. 
In \texttt{Test 1}, the input is a {\tt NumPy} array that matches the focal method's expected type, so the test passes without errors. 
This reveals a key challenge: LLMs tend to imitate the logic of the focal method and produce safe and conventional inputs. Although these inputs are valid, they rarely explore failure-inducing edge cases and are less likely to detect type errors.

To address this limitation, we then prompted the LLM to generate a test explicitly intended to trigger a type error. The resulting \texttt{Test 2} successfully triggers an error by passing a {\tt NumPy} array containing a {\tt NaN} value to the method. 
However, in the real usage of the project, the input is first validated by the method {\tt has\_valid\_item}, which ensures that each element supports certain required magic methods.
Since \texttt{NaN} lacks these methods, it violates the context-aware constraint and actually causes a false positive. 
This motivates the need for extracting context-aware type constraints to guide the test generation process and reduce false positives.

Based on these findings, we further prompted the LLM to generate a test case explicitly aimed at triggering a type error, guided by the type constraint. 
The resulting \texttt{Test 3} shows a common failure mode caused by LLM hallucination. Specifically, the LLM incorrectly assumes that the {\tt TestLocationIndexer} object supports subscript access (i.e., {\tt indexer['x']}), implying the existence of a {\tt \_\_getitem\_\_} method. 
However, this method is not implemented at all, and thus such access is invalid and results in a {\tt TypeError}. This error does not reflect a fault in the focal method but arises from incorrect assumptions in the generated test. 
Such hallucinations further introduce false positives, ultimately undermining the reliability of the testing process. 
This motivates the need for an effective mechanism to identify and refine hallucinated test cases.

To address these challenges, we propose \tool{}, which enhances LLM-based unit test generation with type constraint analysis and self-reflection for precise type error detection. 
Specifically, \tool{} first captures context-aware type constraints via invocation chain analysis and employs an error-seeking agent to propagate constraints backward step-by-step.
Then, the constraints are used to guide the test generation process, aiming to detect type errors. 
Finally, \tool{} applies a reflection mechanism to iteratively refine test generation, reducing false positives in type error detection.
\texttt{Test 4} demonstrates the effectiveness of this technique. It tests the {\tt \_validate\_key} method indirectly via a realistic call chain ({\tt \_\_getitem\_\_ $\rightarrow$ \_getitem\_tuple $\rightarrow$ \_has\_valid\_item $\rightarrow$ \_validate\_key}) and uses a {\tt datetime} object as input. 
The input propagates through multiple internal methods, and ultimately exposes the type error in {\tt \_validate\_key} under the realistic condition.

\section{Approach}

\begin{figure*}[t]
  \centering
  \includegraphics[width=0.85\linewidth]{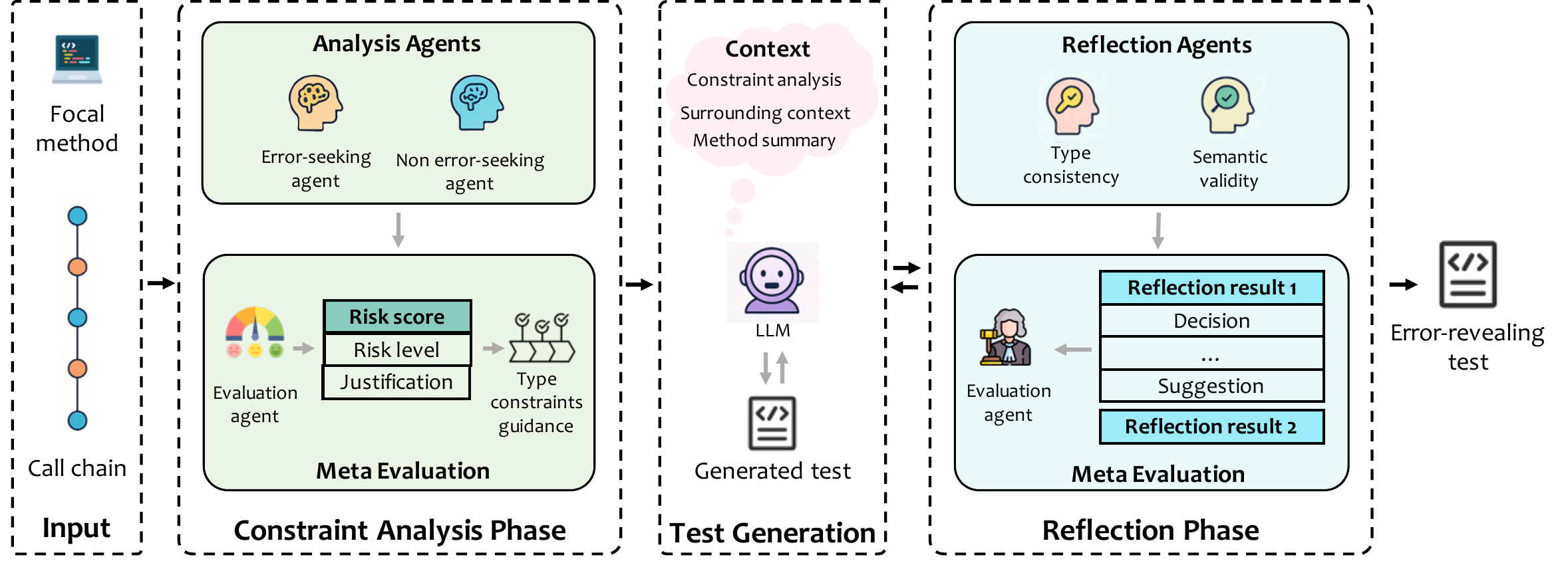}
  \caption{Overview of \tool{}}
  \label{fig:overview}
\end{figure*}

Figure~\ref{fig:overview} provides an overview of \tool{}. 
Given a focal method and its corresponding invocation chain, \tool{} proceeds in three main stages.
First, in the constraint analysis phase, \tool{} infers context-aware type constraints. Specifically, an \erroranalysis{} identifies risky input types and infers the upstream constraints needed to produce them. 
To avoid unrealistic or hallucinated constraints from the LLM, an evaluation agent verifies their feasibility step by step. 
If the inferred constraints are plausible and could lead to errors, the invocation chain is marked as high-risk, and the constraints guide bug-oriented test generation. 
Otherwise, \tool{} uses a \normalanalysis{} strategy to infer likely correct type constraints to ensure testing sufficiency.
Second, in the test generation phase, \tool{} generates tests for the entry method in the invocation chain, guided by the inferred type constraints and its surrounding context (e.g., class fields, related methods).
Finally, in the reflection phase, three specialized agents estimate false positives. A type consistency agent verifies whether the test respects the inferred type constraints. A semantic validity agent checks if the behavior of the test aligns with intended usage. A meta-evaluation agent consolidates these insights and feeds them back to the test generation agent for iterative refinement (if a false positive is estimated).
The key steps are explained in detail in the following sections.

\subsection{Constraint Analysis Phase}

To solve the problem of LLMs failing to generate type-correct and error-revealing unit tests, \tool{} enhances LLM-based test generation by explicitly guiding it with type constraints.
Specifically, \tool{} performs a step-by-step type constraint analysis backward along the invocation chain leading to the focal method to infer context-aware constraints likely to trigger type errors. 
An evaluation agent then estimates the risk of type errors by assessing the feasibility of the inferred constraints at each step along the chain. If the chain is deemed high-risk, the error-revealing constraints are passed to the test generation phase as guidance. Otherwise, \tool{} falls back to a \normalanalysis{} that infers likely correct input types, ensuring test completeness.

\subsubsection{Analysis Agents}
Let the invocation chain be denoted as $\langle F_1, F_2, \ldots, F_n \rangle$, where $F_n$ is the focal method potentially exhibiting a type error. To assist in detecting such errors and generating valid test inputs, LLMs should be guided by type information.
However, due to Python’s lack of explicit type annotations and its flexible type system, analyzing $F_n$ in isolation often provides insufficient context and may lead to incorrect or semantically invalid assumptions. Conversely, analyzing the entire chain in one shot is overly complex and prone to hallucinations due to the extended reasoning path.

\begin{figure}[t]
  \centering
  \includegraphics[width=0.9\linewidth]{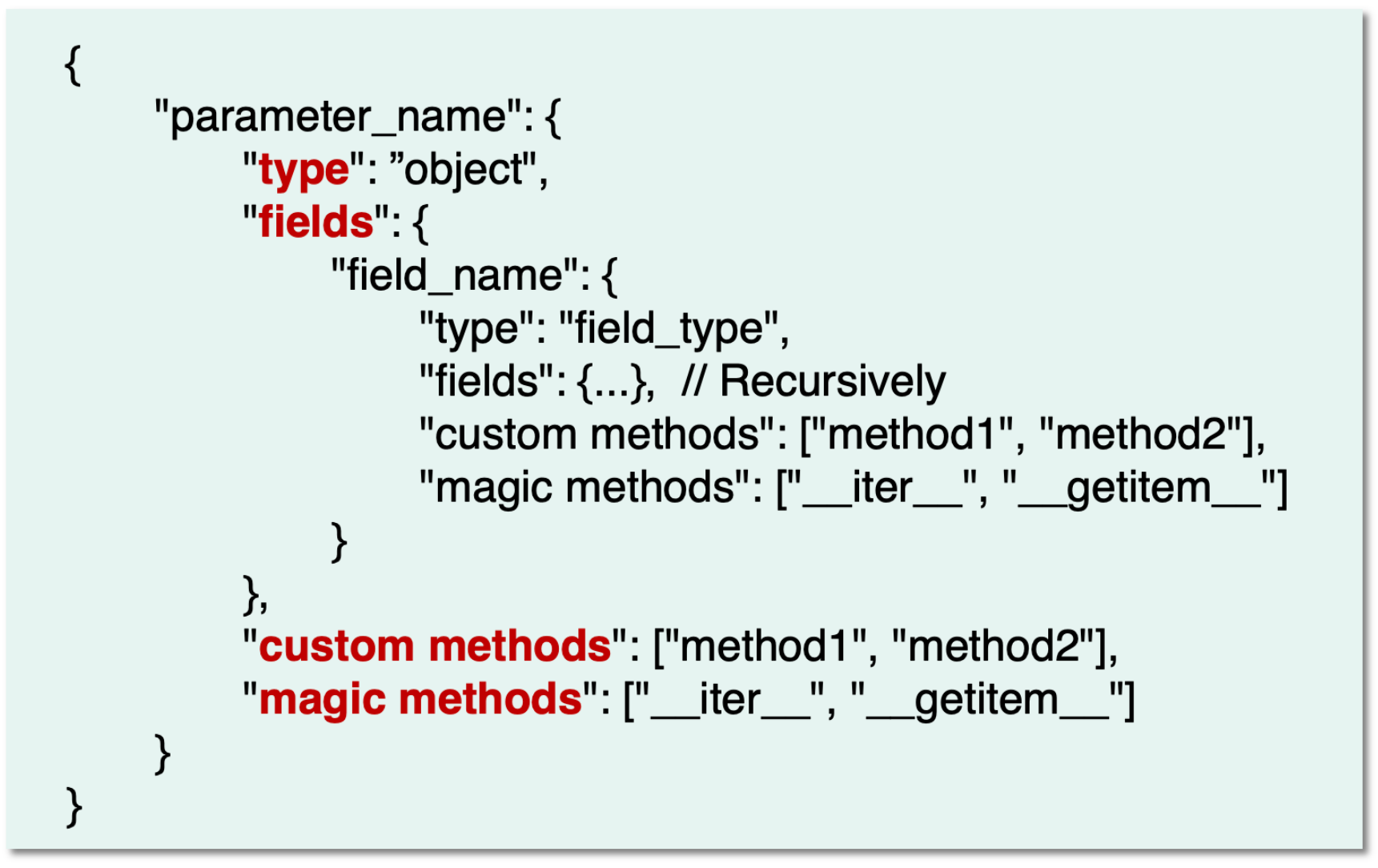}
  \caption{An example of type constraints}
  \label{fig:constraints}
\end{figure}

To this end, \tool{} performs a step-by-step backward analysis, starting from $F_n$ and proceeding to $F_1$. Each step analyzes a single function call $\langle F_i, F_{i-1} \rangle$ and infers the type constraint $P_{i-1}$ for $F_{i-1}$, which describes the expected type constraints of its parameters. However, real-world programs often involve a wide range of types. Representing constraints using concrete types would create an impractically large search space and reduce precision. To mitigate this, \tool{} represents constraints using a structured schema based on Python primitive types and \textit{magic methods}—special methods (e.g., {\tt \_\_getitem\_\_()}, {\tt \_\_iter\_\_()}) that define behaviors for built-in operations (e.g., subscripting, iteration).
Specifically, the constraint associated with each parameter comprises four components:
\begin{itemize}
    \item \textbf{Type}: Indicates whether the parameter is a primitive or a user-defined object.
    \item \textbf{Fields}: Describes the structure and expected type constraints of fields, including nested elements for containers like lists and dictionaries.
    \item \textbf{Custom Methods}: Lists explicitly invoked methods that the parameter should support.
    \item \textbf{Magic Methods}: Lists Python magic methods that the parameter should support.
\end{itemize}
Notably, \tool{} represents each constraint in JSON format, with an example illustrate in Figure~\ref{fig:constraints}.

Since the type constraints are inferred in a backward manner along the invocation chain, each $P_i$ must be compatible with all downstream constraints $P_j$ where $j > i$. That is, at each step, given $\langle F_i, P_i \rangle$, \tool{} infers a corresponding constraint $P_{i-1}$ that ensures $F_{i-1}$ can produce inputs satisfying $P_i$ in its call to $F_i$.
This step-by-step inference process can be formally expressed as:

$$
\langle F_i, P_i, F_{i-1} \rangle \Rightarrow \langle F_i, P_i, F_{i-1}, P_{i-1} \rangle
$$
\tool{} starts by invoking two specialized agents to infer type constraints in $F_{n}$, which serves as the starting point of the entire backward propagation process. Each agent focuses on a different aspect of constraint inference: %

\begin{itemize}
\item Error-seeking agent infers error-triggering constraints $P_n^{\text{trigger}}$ likely to expose type errors in $F_n$.
\item Non-error-seeking agent infers valid-use constraints $P_n^{\text{normal}}$ that allow $F_n$ to execute successfully.
\end{itemize}
These complementary constraints help balance fault detection with test completeness. While the error-seeking agent aims to reveal errors, it may occasionally miss bugs or generate unrealistic constraints. In such cases, the non-error-seeking agent offers a fallback by focusing on typical, safe usage patterns.
Once $F_n$ is analyzed, \tool{} continues type constraint inference for each caller in the chain ($\langle F_n \Rightarrow \dots \Rightarrow F_1$), propagating constraints backward until it reaches $F_1$.

At the end of the backward inference process, \tool{} produces two sequences of type constraints: $\langle P_n^{\text{trigger}}, \dots, P_1^{\text{trigger}} \rangle$ and $\langle P_n^{\text{normal}}, \dots, P_1^{\text{normal}} \rangle$.
These results are then passed to a meta-evaluation phase, where an evaluation agent estimates the likelihood of type errors in the chain and selects the more appropriate constraint set to guide the subsequent test generation process.

\subsubsection{Meta Evaluation}
In the previous step, \tool{} derived two sets of type constraints: one aimed at revealing errors and another representing correct usage. 
Then, \tool{} evaluates the feasibility and risk of the error-revealing constraints using a dedicated evaluation agent. 
This agent validates the constraints and estimates the likelihood that the associated invocation chain could trigger a true type error.
Specifically, the agent produces two outputs:
(1) \textbf{Risk level}—labeled as either \textit{high} or \textit{low}, indicating the estimated probability of encountering a type error;
(2) \textbf{Justification}—a concise explanation summarizing the rationale behind the risk assessment, including which part of the chain contributes most to the potential error.

Ideally, invocation chains with plausible error-revealing constraints are classified as high-risk, and these constraints will be directly used to guide bug-oriented test generation. 
Otherwise, \tool{} falls back to the valid-usage constraints inferred by \normalanalysis{} to ensure test completeness.

\subsection{Test Generation Phase}
\tool{} treats \textit{the entry method} of the previously analyzed invocation chain as the method to be tested. This design choice ensures that the generated test cases reflect realistic usage scenarios (i.e., the entry point of a project or module would be invoked in practice). 
The goal of the test generation phase is to create test cases guided by the type constraints and invocation chain, along with the contextual information of the \textit{entry method}.

\subsubsection{Context Collection}
Prior work on LLM-based test generation has shown that incorporating contextual information beyond the focal method substantially improves the quality of generated tests~\cite{chattester, ase_study, TELPA}. 
Motivated by this insight, \tool{} constructs comprehensive context to better support test generation. It collects two types of context: \textit{cross-file} and \textit{intra-file} context. 
The cross-file context includes the full invocation chain analyzed during type constraint analysis, along with the inferred constraints themselves.
The intra-file context is gathered by parsing the file hosting the entry method and extracting relevant program elements, specifically including all import statements, global fields, class definitions, and method definitions. If the \textit{entry method} belongs to a class, \tool{} also collects the class constructor and any class methods that are directly invoked by the \textit{entry method}.
The context information is then used to guide the LLM during test generation.

\subsubsection{Test Generation Process}
To guide the LLM with the previously extracted context, \tool{}
begins by inserting the cross-file context, including the invocation chain and its inferred type constraints, into the test generation agent’s memory as the chat history. Specifically, each entry (i.e., a round of conversation with the LLM) in this history corresponds to a step from the earlier constraint analysis phase, capturing the type constraint for a single call in the invocation chain. This equips the agent with essential background knowledge on the method’s broader usage context.
Next, inspired by the success of Chain-of-Thought (CoT) prompting~\cite{cot}, \tool{} employs a two-stage CoT strategy. In the first stage, it formats the intra-file context to reflect the original source structure, appending the entry method at the end. The LLM is then prompted to summarize the method’s functionality according to the context, improving its contextual awareness of the method's semantics. Both the intra-file context (included in the prompt used to instruct the LLM) and the generated functionality summary are also retained in memory as a round of conversation in the chat history.
In the second stage, the LLM is prompted to generate unit tests using the stored cross-file context, intra-file context, and method summary. This combination serves as rich guidance, enabling the model to produce more targeted and meaningful test cases.

For each generated test case, \tool{} first removes all assertions, as they are unnecessary for detecting type errors. \tool{} then executes the test. If the test fails without raising a type error, \tool{} invokes a self-debugging step~\cite{self-debug}, prompting the LLM to revise the test based on the error message and the original code. 
If the revised test still fails without raising a type error, \tool{} discards it. 
Otherwise, if a type error is raised, \tool{} proceeds to the reflection phase to determine whether the error is a true or false positive.
 
\subsection{Reflection Phase}
\begin{figure*}[t]
  \centering
  \includegraphics[width=0.8\linewidth]{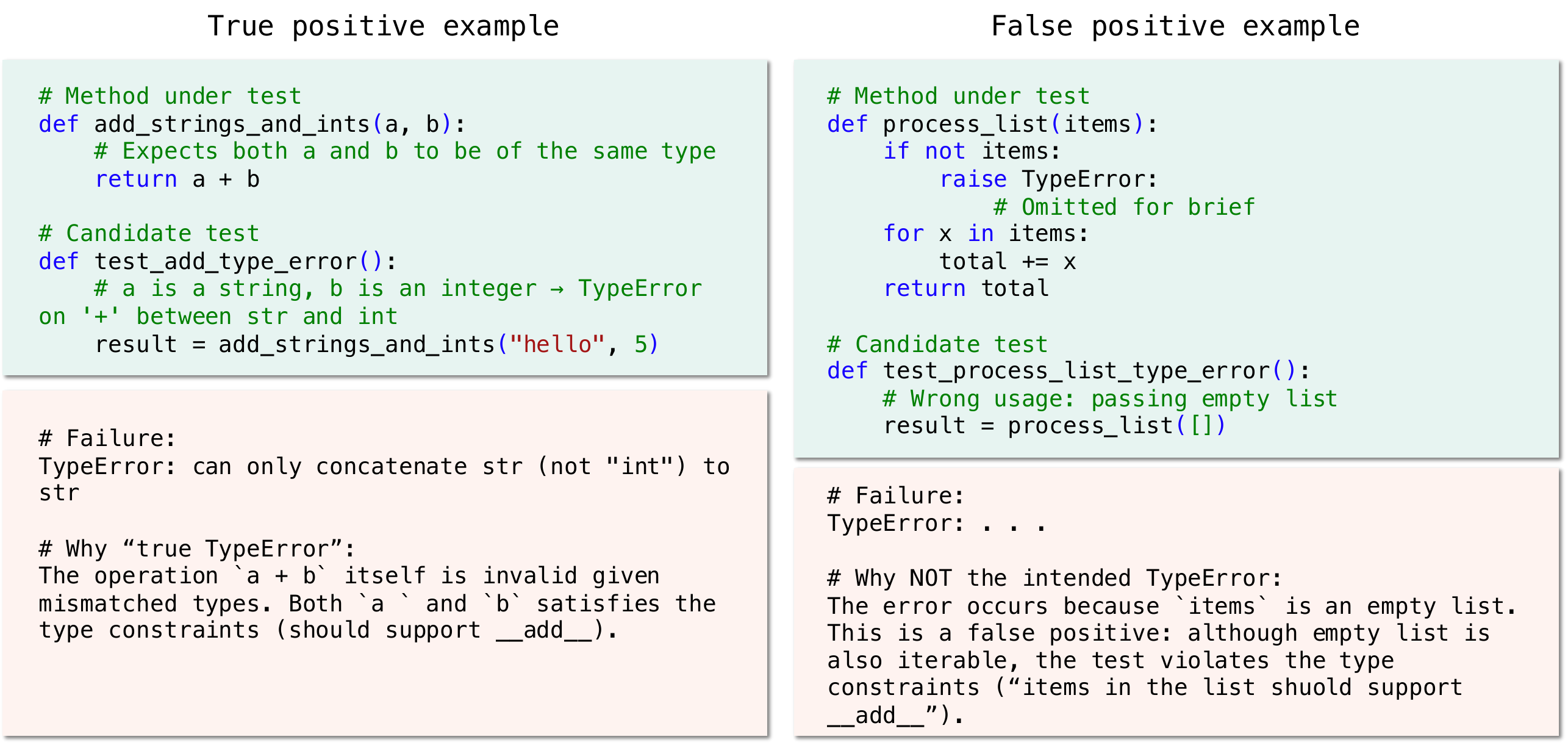}
  \caption{Few-shot examples for type consistency agent}
  \label{fig:type_example}
\end{figure*}

Existing researches have pointed out that LLMs are prone to hallucinations, which can lead to the generation of invalid or misleading unit tests~\cite{chattester, symprompt, ase_study}. Therefore, when an LLM-generated test case triggers a type error, it is essential to determine whether the error genuinely reflects a type misuse or is simply a result of an ill-formed test (e.g., using an invalid parameter type) to minimize false positives. To this end, \tool{} employs three specialized agents: two reflection agents (one for type consistency checking and another for semantic validity checking) and a meta-evaluation agent. Together, these agents validate generated failures and iteratively refine the test generation process.

\subsubsection{Reflection Agents}

As discussed in Section~\ref{sec:intro}, a valid type error test case must satisfy two conditions: 
(1) type consistency—the test input should satisfy the inferred type constraints, and 
(2) semantic validity—the error should arise from a meaningful usage scenario rather than unrealistic scenarios or unrelated logic issues.
To enforce these criteria, \tool{} employs two specialized reflection agents:
\begin{itemize}
    \item \textbf{Type Consistency Agent}: Checks whether the test inputs align with the inferred type constraints.
    \item \textbf{Semantic Validity Agent}: Verifies that the test reflects a realistic use case and that the error genuinely stems from a true type misuse.
\end{itemize}
Each agent is provided with the inferred type constraints used to generate the test case, the content of the invocation chain, the generated test case, and its execution output, including error messages and stack traces.
To compensate for LLMs’ limited domain knowledge in diagnosing type-related false positives, \tool{} adopts a few-shot learning approach. Each agent is guided by two curated examples—one illustrating a true type error and another showing a false positive due to invalid input. Figure~\ref{fig:type_example} presents the examples for the type consistency agent. Due to space limitation, we put other examples used on our project homepage~\cite{homepage}.

Both agents produce outputs with four components:
\begin{itemize}
    \item \textbf{Decision}: Classifies the test as a true or false positive.
    \item \textbf{Confidence}: Indicates certainty (high, medium, low), which will be used in subsequent meta-evaluation.
    \item \textbf{Rationale}: Summarizes the rationale behind the decision.
    \item \textbf{Suggestions}: If the test is deemed false positive, offers guidance for fixing the test.
\end{itemize}
The results from the two reflection agents are then forwarded to a third agent, i.e., the evaluation agent, which acts as an arbiter and makes the final decision.

\subsubsection{Meta Evaluation}
The meta-evaluation agent is responsible for aggregating the outputs of the reflection agents to make a final decision. 
It uses an LLM-based weighted voting strategy that considers structured outputs from the type consistency and semantic validity agents, along with the inferred type constraints and the associated invocation chain. Based on this information, it classifies the test case as either a true positive or a false positive.
If the test is deemed a true positive, the test case is retained. Otherwise, the agent synthesizes an explanation along with actionable suggestions based on the rationales from both reflection agents. These suggestions are then fed back to the test generation agent as iterative feedback.
This feedback loop allows the test generation agent to refine its output, improving its ability to produce truly bug-revealing test cases while minimizing false positives.
If, after refinement, the LLM still does not trigger a true type error, \tool{} proceeds to test the next method.

\section{Evaluation Design}
\subsection{Research Questions}
To evaluate the effectiveness of \tool{}, we formulate the following research questions (RQs).
\begin{itemize}
    \item \textbf{RQ1: To what extent can \tool{} detect type errors compared to state-of-the-art techniques?} 
    The goal of \tool{} is to generate unit tests for detecting python type errors, and this RQ investigates its ability in this regard. 
    
    \item \textbf{RQ2: How does each main component of \tool{} contribute to its overall effectiveness?} This RQ motivates an ablation study to evaluate the individual impact of \tool{}'s core components on its effectiveness.
    
    \item \textbf{RQ3: Can \tool{} uncover previously unknown type errors in real-world Python projects?} This RQ assesses the practical applicability of \tool{} by evaluating its ability to discover new type-related bugs in real-world Python codebases.
\end{itemize}

\subsection{Subjects}
\label{sec:subject}
To answer RQ1 and RQ2, we adopted two widely-used benchmarks, i.e., BugsInPy~\cite{bugsinpy} and TypeBugs~\cite{pyter}, following prior work~\cite{pyter, typefix, pyinder}.
Each benchmark contains a collection of real-world Python type-related bugs along with their corresponding fixed versions. 
To avoid duplication, we removed overlapping bugs between the two benchmarks following the practice of the existing work~\cite{pyinder, d3, chen2023toward}.

We then attempted to replicate those type-related bugs. However, this replication process presented several challenges. 
For instance, some required third-party libraries had been removed from PyPI and could no longer be installed (e.g., the {\tt codecov} package required by several bugs in the {\tt core} project\footnote{\url{https://github.com/home-assistant/core/issues/91283}}). 
Additionally, many benchmark-listed dependencies conflicted with each other, a problem also noted in various related GitHub issues\footnote{\url{https://github.com/kupl/PyTER/issues/1}}\footnote{\url{https://github.com/kupl/PyTER/issues/2}}\footnote{\url{https://github.com/JohnnyPeng18/TypeFix/issues/1}}. 
We manually resolved these dependency conflicts to the best of our ability.
As a result, we successfully replicated 69 real-world type errors across 16 open-source projects in the two benchmarks, with codebases ranging from 3K to 316K lines of code.

For each bug, we extracted the method responsible for triggering the type error, referred to as the buggy focal method. These methods form the basis for evaluating each technique’s effectiveness in detecting Python type errors.
We also extracted the corresponding fixed versions, referred to as non-buggy focal methods, to assess whether a unit test generation technique can avoid producing false positives on correct code.
In total, we obtained 138 methods, including 69 buggy ones and 69 non-buggy ones.

To answer RQ3, we applied \tech{} to the latest versions of six popular open-source Python projects to evaluate its effectiveness in detecting previously unknown type errors.
Project selection was guided by three criteria:
(1) We excluded repositories that primarily serve as educational resources, tutorials, or textbook materials, as they do not reflect production-level complexity.
(2) We included only actively maintained projects with an average commit interval of less than one week.
(3) We selected projects with comprehensive documentation and clear setup instructions to ensure compatibility with our experimental environment.
We examined top-ranked Python repositories on GitHub (sorted by star count) and retained 50 projects that satisfied all three criteria. 
Then, considering evaluation costs, we sampled six for this experiment to avoid subjective bias.
They are kivy~\cite{kivy}, langchain~\cite{langchain}, luigi~\cite{luigi}, pwntools~\cite{pwntools}, scipy~\cite{scipy}, and scrapy~\cite{scrapy}.
\ins{
Note that we did not simply select the top-starred projects, as these are often dominated by AI libraries with highly similar code patterns and error types. Instead, our sampling strategy prioritized domain diversity and evaluation reliability.
}

\subsection{Compared Techniques}
\tool{} aims to generate effective unit tests for precise type error detection, and thus we adopted both the state-of-the-art type error detection technique and the state-of-the-art LLM-based unit test generation techniques as our baselines:

\begin{itemize}
    \item \textbf{\pyinder{}}~\cite{pyinder}: It is a static type-error detection tool for Python, which incorporates four key features identified through manual investigation for type error detection.
    
    \item  \textbf{\tester{}}~\cite{chattester}: The first technique to leverage LLMs for unit test generation. It prompts the LLM with the focal method and its context to generate tests.
    
    \item \textbf{\sym{}}~\cite{symprompt}: It prompts LLMs to generate one test per execution path of the focal method, aiming to improve code coverage by encouraging path diversity.
    
    \item \textbf{\hits{}}~\cite{hits}: It first leverages LLMs to decompose complex methods into smaller slices and then guides LLMs to generate tests for each slice independently, aiming to improve code coverage.
\end{itemize}

We did not include the traditional Python unit test generation tool (i.e, Pynguin~\cite{pynguin}) for comparison. 
This is because
(1) Pynguin fails to run on many projects in the two benchmarks due to its limited support for Python versions. 
(2) Many existing studies have demonstrated that LLM-based unit test generation outperforms the traditional Pynguin~\cite{symprompt,codamosa,TELPA}.

\subsection{Measurements}
\label{sec:metric}
These studied techniques differ in output: test generation techniques produce tests that may trigger runtime type errors, while \pyinder{} raises static alarms.
Therefore, we unify the evaluation by considering a type error as ``reported'' for a given focal method if a test triggers a type error or a static tool raises an alarm.

\subsubsection{Outcomes on Buggy Methods}
To evaluate the effectiveness in triggering type errors, we ran each test generation technique on the buggy method and executed the generated tests on both the buggy and the corresponding fixed versions. 
For \pyinder{}, we applied it to both the buggy and fixed methods.
Following the existing definition~\cite{toga}, we categorize the outcomes of each technique as follows:

\begin{itemize}
\item True Positive for bug detection (TP$_{\textit{bug}}$): 
The technique reports a type error only on the buggy version, but not on the corresponding fixed version.

\item False Positive for bug detection (FP$_{\textit{bug}}$): The technique reports a type error on both the buggy and fixed versions.

\item False Negative for bug detection (FN$_{\textit{bug}}$): The technique fails to report a type error on the buggy version.

\end{itemize}
Since all target methods under this setting are buggy (i.e., positive samples), there are no true negatives.

\subsubsection{Outcomes on Non-buggy Methods}
To investigate whether \tech{} can avoid producing false positives on correct code, we ran each technique on the fixed version. 
The possible outcomes include:
\begin{itemize}
\item False Positive on non-buggy methods (FP$_{\textit{nonbug}}$): The technique incorrectly reports a type error on the non-buggy version.

\item True Negative on non-buggy methods (TN$_{\textit{nonbug}}$): The technique correctly does not report any error on the non-buggy version.
\end{itemize}
All target methods under this setting are non-buggy (i.e., negative samples), and thus there are no true positives and false negatives.

\subsubsection{Metric Calculation}
Based on these outcomes, we measured the effectiveness of each technique using the following metrics:

\textbf{Accuracy} measures the proportion of correct identifications, i.e., detecting type errors in buggy methods while correctly not detecting errors in non-buggy methods. It is calculated as:
$\frac{ \text{TP}_\textit{bug} +  \text{TN}_\textit{nonbug}}{ \text{TP}_\textit{bug} +  \text{FP}_\textit{bug} +  \text{FP}_\textit{nonbug} +  \text{TN}_\textit{nonbug} +  \text{FN}_\textit{bug}}$.

\textbf{Precision} measures the proportion of true bugs among all samples identified as bugs:
$\frac{ \text{TP}_\textit{bug}}{ \text{TP}_\textit{bug} +  \text{FP}_\textit{bug} +  \text{FP}_\textit{nonbug}}$.

\textbf{Recall} measures the proportion of true bugs correctly identified out of all true bugs: 
$\frac{ \text{TP}_\textit{bug}}{ \text{TP}_\textit{bug} +  \text{FN}_\textit{bug}}$.

\textbf{F1-score} is the harmonic mean of Precision and Recall, providing a balanced measure that accounts for both false positives and false negatives:
$\frac{2 \times \text{Precision} \times \text{Recall}}{\text{Precision} + \text{Recall}}$.

\subsection{Implementation and Environment}
We implemented \tool{} in Python, utilizing Jarvis~\cite{jarvis} and tree-sitter~\cite{tree-sitter} for call chain extraction. 
\ins{
Jarvis combines flow-sensitive intra-procedural analysis and inter-procedural analysis to infer types and handle dynamic dispatch. Its type inference engine approximates runtime variable types to construct a receiver-type-aware call graph, allowing RTED to retrieve a reasonably sound set of call chains for focal methods.
}
As the underlying LLM, we used DeepSeek-V3~\cite{deepseek} via its API to power all agents within \tool{}.
For \pyinder{}, we directly leveraged the released implementation~\cite{pyinder_artifact}.
For \tester{} (which is originally designed for Java), we adapted it to Python based on its publicly available implementation. For \sym{} and \hits{}, due to the lack of released code, we re-implemented them based on the descriptions in their respective papers. To ensure a fair comparison, we used the same DeepSeek-V3 model~\cite{deepseek} as the underlying LLM for \tester{}, \sym{}, and \hits{}.
For all focal methods in the benchmarks, we configured each technique to generate one test file per focal method. 
For \tool{}, one representative call chain \ins{(i.e., the shortest)} is sampled per method to guide test generation. 
\ins{This design reduces prompt ambiguity and ensures the input remains within the LLM’s context window, since \tech{} provides all function implementations along the selected call chain as context.}
This also ensures that the resulting test suites are of comparable scale across techniques, enabling a fair comparison.
We executed all generated tests using each project's original testing framework, primarily pytest~\cite{pytest} or unittest~\cite{unittest}. 
All experiments were conducted on a workstation running Ubuntu 20.04, equipped with a 128-core CPU, 504 GB of RAM, and 4 NVIDIA A800 GPUs.

\section{Results and Analysis}
\label{sec:rq1}
\subsection{RQ1: Effectiveness on Type Error Detection}
\subsubsection{Process}
For \pyinder{}, we collected its reported alarms. For test generation approaches, we generated and executed test suites to observe whether any type errors were triggered during runtime. Each technique's output was then mapped to one of several possible outcomes introduced in Section~\ref{sec:metric} based on whether it reported a type error for the buggy or fixed version of a method.

\subsubsection{Results}

\begin{table}[t]
\renewcommand{\arraystretch}{1.2}
\caption{Comparison among \tool{}, \pyinder{}, \tester{}, \sym{}, and \hits{}}
\centering
\label{tab:rq1}
\resizebox{0.95\linewidth}{!}
{
\begin{threeparttable}
\begin{tabular}{l|rrrr|rrrr}
\toprule
\multicolumn{1}{c|}{\multirow{2}{*}{\textbf{App.}}} & \multicolumn{4}{c|}{\textbf{BugsInPy}} & \multicolumn{4}{c}{\textbf{TypeBugs}} \\
\multicolumn{1}{c|}{}                               & \multicolumn{1}{c}{\textbf{P}} & \multicolumn{1}{c}{\textbf{R}} & \multicolumn{1}{c}{\textbf{F1}} & \multicolumn{1}{c|}{\textbf{Acc}} & \multicolumn{1}{c}{\textbf{P}} & \multicolumn{1}{c}{\textbf{R}} & \multicolumn{1}{c}{\textbf{F1}} & \multicolumn{1}{c}{\textbf{Acc}} \\
\midrule
\pyinder{} & 0.25&	0.11&	0.15&	0.45&		0.14&	0.06&	0.08&	0.44 \\
\tester{}                                         & 0.25                                   & 0.13                                & 0.17                            & 0.50                                  & 0.29                                   & 0.11                                & 0.16                            & 0.47                                  \\
\sym{}                                          & 0.09                                   & 0.04                                & 0.06                            & 0.43                                  & 0.33                                   & 0.05                                & 0.09                            & 0.50                                  \\
\hits{}                                               & 0.11                                   & 0.09                                & 0.10                            & 0.40                                  & 0.34                                   & 0.31                                & 0.33                            & 0.47                                  \\
\midrule
\tool{}                                                & \textbf{0.78}                                   & \textbf{0.50}                                & \textbf{0.61}                            & \textbf{0.70}                                  & \textbf{0.69}                                   & \textbf{0.54}                                & \textbf{0.61}                            & \textbf{0.67}          \\
\bottomrule
\end{tabular}
\end{threeparttable}
}
\end{table}

\begin{figure}[t]
  \centering
  \includegraphics[width=0.9\linewidth]{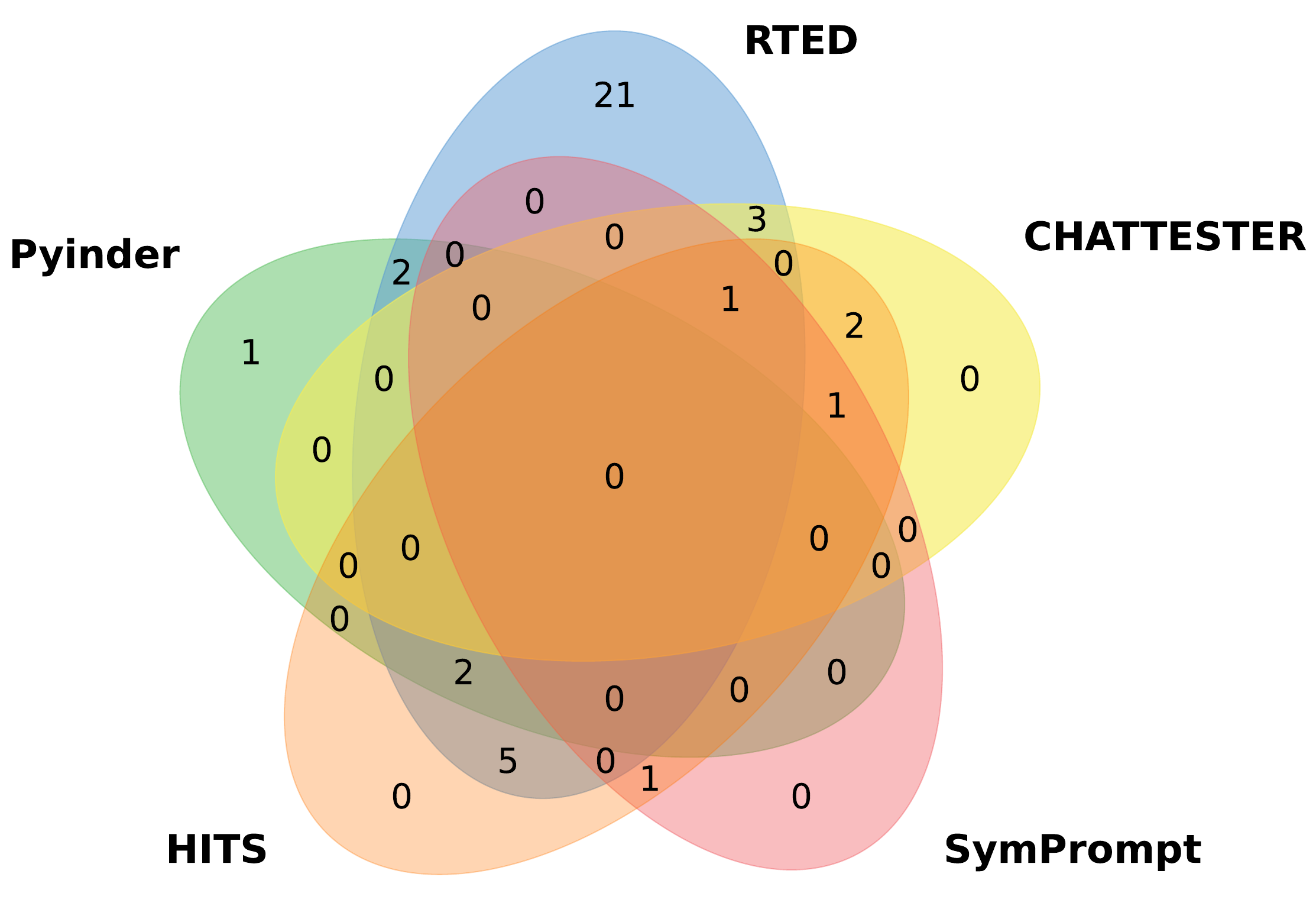}
  \caption{Overlap of bug detection}
  \label{fig:rq1_venn}
  \vspace{-1em}
\end{figure}

Table~\ref{tab:rq1} presents the overall results in terms of precision (denoted as \textbf{P}), recall (denoted as \textbf{R}), F1-score (denoted as \textbf{F1}), and accuracy (denoted as \textbf{A}).
From the table, \tool{} consistently outperforms all baselines, including the static analyzer \pyinder{} and dynamic test generation techniques, across both benchmarks. 
In terms of accuracy, \tool{} achieves 0.70 on BugsInPy and 0.67 on TypeBugs, significantly outperforming the best-performing baselines: 0.50 by \tester{} on BugsInPy and 0.50 by \sym{} on TypeBugs, achieving improvements of 40\% and 34\%, respectively.
For F1-score, \tool{} reaches 0.61 on both datasets, while the best baselines, \tester{} (0.17 on BugsInPy) and \hits{} (0.33 on TypeBugs), fall far behind. This demonstrates that \tool{} not only detects more bugs but also maintains strong precision, avoiding the excessive false positives that undermine many existing techniques.

Figure~\ref{fig:rq1_venn} shows the overlap in bugs detected by each technique. From the figure, \tool{} successfully detects 34 bugs, while \pyinder{}, \tester{}, \sym{}, and \hits{} detect only 5, 7, 3, and 12 bugs, respectively. 
Moreover, to detect those bugs, \pyinder{}, \tester{}, \sym{}, and \hits{} produces 21, 19, 14, 35 false positives respectively, while \tool{} only produces 13 false positives, 
achieving an improvement of 173.9\%$\sim$245.9\% in precision. 
Notably, \tool{} not only covers nearly all bugs found by others but also identifies the largest number of unique bugs with fewest false positives, underscoring its practical effectiveness.

Note that \pyinder{}, despite being tailored for type error detection, performs worse than the test generation approaches in some cases. This is partly due to its inherent limitations: certain bugs in the benchmarks require runtime information or involve third-party libraries, which static analysis alone cannot effectively handle.
For more direct comparison, we also evaluate the performance of \pyinder{} and \tool{} on the 40 focal methods supported by \pyinder{}, including 20 buggy ones and 20 non-buggy ones. 
On those methods, \pyinder{} detects only 5 type errors and raises 21 false positives. In contrast, \tool{} detects 7 bugs while producing just 2 false positives, highlighting its superior recall and precision, both of which are critical for practical usability.

\begin{tcolorbox}[colback=gray!5]
\textbf{RQ1 Summary}: \tool{} outperforms both static and dynamic baselines in detecting Python type errors, achieving the highest precision, recall, F1-score, and accuracy across two benchmarks. 
That is, it detects the most bugs, including many unique bugs, while maintaining low false positive rates. These results affirm \tool{}’s effectiveness for Python type error detection.
\end{tcolorbox}

\subsection{RQ2: Ablation Study}

\subsubsection{Process}
In this RQ, we investigate the contributions of the two key components in \tool{} (e.g., the constraint analysis phase and the reflection phase). To isolate their effects, we construct two variants of \tool{}:
\begin{itemize}
    \item \textbf{\noanalysis{}}, which removes the constraint analysis component. The LLM generates tests without guidance about type constraints, but still retains the reflection mechanism to iterate based on feedback from test execution.
    \item \textbf{\noreflection{}}, which removes the reflection component. The LLM is still guided by the results of type constraint analysis to generate unit tests. However, no reflection is applied if a generated test triggers a type error.
\end{itemize}
We apply both variants to all focal methods using the same setup as RQ1 and analyze their effectiveness.

\subsubsection{Results}

\begin{table}[t]
\renewcommand{\arraystretch}{1.2}
\caption{Comparison between \tool{} and its variants}
\centering
\label{tab:rq2}
\resizebox{0.95\linewidth}{!}
{
\begin{threeparttable}
\begin{tabular}{l|rrrr|rrrr}
\toprule
\multicolumn{1}{c|}{\multirow{2}{*}{\textbf{App.}}} & \multicolumn{4}{c|}{\textbf{BugsInPy}} & \multicolumn{4}{c}{\textbf{TypeBugs}} \\
\multicolumn{1}{c|}{}                               & \multicolumn{1}{c}{\textbf{P}} & \multicolumn{1}{c}{\textbf{R}} & \multicolumn{1}{c}{\textbf{F1}} & \multicolumn{1}{c|}{\textbf{Acc}} & \multicolumn{1}{c}{\textbf{P}} & \multicolumn{1}{c}{\textbf{R}} & \multicolumn{1}{c}{\textbf{F1}} & \multicolumn{1}{c}{\textbf{Acc}} \\
\midrule
\noanalysis{}                   & 0.64      & 0.25   & 0.36 & 0.58     & 0.58      & 0.33   & 0.42 & 0.62     \\
\noreflection{} & 0.48      & 0.58   & 0.53 & 0.58     & 0.51      & 0.59   & 0.55 & 0.60     \\
\midrule
\tool{}                         & \textbf{0.78}      & \textbf{0.50}   & \textbf{0.61} & \textbf{0.70}     & \textbf{0.69}      & \textbf{0.54}   & \textbf{0.61} & \textbf{0.67}     \\
\bottomrule
\end{tabular}
\end{threeparttable}
}
\vspace{-2em}
\end{table}

Table~\ref{tab:rq2} shows that both ablations lead to noticeable performance drops compared to the full \tool{}, confirming the importance of each component.
Specifically, without explicit type constraint guidance, the LLM struggles to generate tests that expose type errors. This is reflected in drastically reduced recall (0.25 on BugsInPy and 0.33 on TypeBugs) and a correspondingly low F1-score (0.36 on BugsInPy and 0.42 on TypeBugs). Despite a somewhat decent precision (0.64 on BugsInPy and 0.58 on TypeBugs), this variant misses a substantial number of actual type errors, indicating that LLMs alone are insufficiently aware of subtle type issues in the absence of type constraints guidance. The lowered accuracy (0.58/0.62) further shows that this variant misidentifies many buggy methods as safe.

In contrast, when reflection is removed, recall remains relatively high (0.58/0.59), showing that initial type-guided prompts are often effective at detecting bugs. However, precision drops sharply (to 0.48/0.51), meaning the tests also trigger many spurious errors. Without reflection, the LLM cannot refine or validate the generated tests, resulting in a higher rate of false positives. The lower accuracy (0.58/0.60) also reflects this unreliability in distinguishing true bugs.

The full version of \tool{}, combining both type constraints and reflective refinement, achieves the best performance across all metrics. Notably, it achieves the highest F1-score (0.61 on both datasets) and accuracy (0.70/0.67). Compared to \noanalysis{}/\noreflection{}, it improves F1 by 56\%/13\%, respectively. This shows that the synergy between the two components is critical: constraint analysis narrows the test generation space toward effective type error detection, while reflection eliminates false positives, enhancing reliability.

\begin{tcolorbox}[colback=gray!5]
\textbf{RQ2 Summary}: Both type constraint analysis and reflection are essential to \tool{}’s effectiveness. Type analysis steers test generation toward likely type errors, while reflection filters out false positives. Their combination enables \tool{} to achieve a strong balance of precision, recall, and accuracy.
\end{tcolorbox}

\subsection{RQ3: Detecting New Type Errors}
\subsubsection{Process}
In this RQ, we investigate \tool{}'s ability to detect previously unknown type errors in real-world, large-scale Python projects. 
As introduced in Section~\ref{sec:subject}, we selected six open-source Python projects for evaluation. We first used Jarvis~\cite{jarvis} to extract call graphs from each project. Then, we identified top-level entry points, and extracted downstream call chains. \tool{} was then applied to generate and execute unit tests for each of the call chains. 
For comparison, we applied three baseline test generation techniques to all testable public methods in each project. \pyinder{} was run on entire codebases as its whole-program analysis requires. All reported errors were manually verified, and potential type errors were reported to developers for confirmation.

\subsubsection{Results}

\begin{table}[t]
\caption{Comparison among \tool{}, \pyinder{}, \tester{}, \sym{}, and \hits{} in detecting bugs (\textbf{Chat} represents \tester{} and \textbf{Sym} represents \sym{})}
\centering
\label{tab:rq3}
\resizebox{0.95\linewidth}{!}
{
\begin{threeparttable}
\begin{tabular}{l|ccccc}
\toprule
\multicolumn{1}{c|}{\textbf{Bug}} & \multicolumn{1}{c}{\textbf{\pyinder{}}} & \multicolumn{1}{c}{\textbf{
Chat}} & \multicolumn{1}{c}{\textbf{Sym}} & \multicolumn{1}{c}{\textbf{\hits{}}} & \multicolumn{1}{c}{\textbf{\tech{}}} \\
\midrule
kivy-1      &   \cmark   & \cmark   & \xmark    & \cmark   & \cmark    \\ 
kivy-2      &   \xmark   & \xmark   & \xmark    & \cmark   & \cmark    \\
kivy-3      &   \cmark   & \xmark   & \xmark    & \xmark   & \cmark    \\
kivy-4      &   \xmark   & \xmark   & \xmark    & \xmark   & \cmark    \\
langchain-1 &   \xmark   & \xmark   & \xmark    & \xmark   & \cmark    \\
langchain-2 &   \xmark   & \xmark   & \xmark    & \xmark   & \cmark    \\
langchain-3 &   \xmark   & \xmark   & \cmark    & \cmark   & \cmark    \\
luigi-1     &   \xmark   & \xmark   & \xmark    & \xmark   & \cmark    \\
luigi-2     &   \xmark   & \cmark   & \cmark    & \cmark   & \cmark    \\  
luigi-3     &   \cmark   & \cmark   & \cmark    & \cmark   & \cmark    \\   
luigi-4     &   \cmark   & \xmark   & \xmark    & \xmark   & \cmark    \\  
pwntools-1  &   \cmark   & \xmark   & \xmark    & \xmark   & \cmark    \\  
scipy-1     &   \cmark   & \xmark   & \xmark    & \xmark   & \cmark    \\  
scrapy-1    &   \cmark   & \cmark   & \xmark    & \xmark   & \cmark    \\ 
scrapy-2    &   \xmark   & \xmark   & \xmark    & \cmark   & \cmark    \\       
\midrule
Total       &     7      &   4      &   3       &  6       &    15     \\
\bottomrule
\end{tabular}
\end{threeparttable}
}
\end{table}

Table~\ref{tab:rq3} presents the detailed results. 
\tool{} successfully detected \found{} previously unknown type errors, outperforming all other techniques. 
Additionally, three bugs (i.e., kivy-3, langchain-3, and scrapy-2) were identified as duplicates of existing reports filed by other users but had not yet been fixed (indicating already known bugs). This suggests that \tool{} is effective at discovering bugs that align with realistic usage scenarios encountered by actual users.
In comparison, excluding duplicates, \pyinder{} detected 6 unknown bugs, \hits{} detected 4, while \tester{} and \sym{} detected only 4 and 1 unknown bugs, respectively. 
Among the \found{} unknown bugs reported by \tool{}, four have been confirmed or fixed by the developers.
This result highlights \tool{}'s superior capability in exposing type-related issues in real-world codebases.

\lstdefinestyle{mystyle}{
    numbers=left,                   %
    numberstyle=\tiny\color{gray},   %
    basicstyle=\ttfamily\scriptsize,
    lineskip=0.8pt,                   %
    keywordstyle=\color{magenta},    %
    commentstyle=\color{violet!60!gray}, %
    stringstyle=\color{blue},         %
    backgroundcolor=\color{white}, %
    showstringspaces=false,          %
    xleftmargin=1.5em,               %
    xrightmargin=1.5em,              %
    frame=single,                    %
    breaklines=true,                 %
    moredelim=**[is][\color{red!50}]{`}{`}, %
    escapeinside=``,                 %
    columns==fullflexible,
}

\lstset{style=mystyle}

\begin{lstlisting}[language=Python, caption={A simplified version of pwntools-1}, label={listing:bug_example}]
getattr(input_stream, 'buffer', input_stream).readline(_size).rstrip(b'\n') 
\end{lstlisting}

Our manual analysis confirms that \tool{} effectively detects bugs involving subtle type constraints that are often missed by conventional test generation techniques. As shown in Listing~\ref{listing:bug_example}, the code calls \texttt{rstrip(b"\textbackslash n")} on the result of \texttt{readline()}, assuming it returns a byte string. However, if \texttt{readline()} instead returns a regular string, a \texttt{TypeError} occurs because \texttt{str.rstrip()} does not accept a \texttt{bytes} argument.
To trigger this bug, the test input must first \textit{satisfy a reachability constraint}: it must implement a \texttt{readline()} method whose return value supports \texttt{rstrip()}. Otherwise, an \texttt{AttributeError} would be raised before reaching the buggy line. 
Then, the input should not return the expected \texttt{bytes} for {\tt readline()}, thereby exposing the type mismatch.
\tool{} is able to detect this issue by first inferring a type that could trigger an error (e.g., {\tt str}). It then propagates backward to supplement additional constraints (i.e., supports {\tt readline()}), and finally uses a {\tt StringIO} object as input to expose the bug.
In contrast, all other evaluated test generation tools fail to detect this issue.

\begin{tcolorbox}[colback=gray!5]
\textbf{RQ3 Summary}: \tool{} demonstrates strong practical effectiveness in detecting previously unknown type errors in actively maintained open-source projects, highlighting its potential to enhance the reliability of modern Python software.
\end{tcolorbox}

\section{Discussion}

\begin{table}[t]
\caption{Generalizability of \tech{} on different LLMs}
\label{tab:different_llm}
\renewcommand{\arraystretch}{1.2}
\centering
\resizebox{0.95\linewidth}{!}
{
\begin{threeparttable}
\begin{tabular}{l|rrrr|rrrr}
\toprule
\multicolumn{1}{c|}{\multirow{2}{*}{\textbf{App.}}} & \multicolumn{4}{c|}{\textbf{BugsInPy}} & \multicolumn{4}{c}{\textbf{TypeBugs}} \\
\multicolumn{1}{c|}{}                               & \multicolumn{1}{c}{\textbf{P}} & \multicolumn{1}{c}{\textbf{R}} & \multicolumn{1}{c}{\textbf{F1}} & \multicolumn{1}{c|}{\textbf{Acc}} & \multicolumn{1}{c}{\textbf{P}} & \multicolumn{1}{c}{\textbf{R}} & \multicolumn{1}{c}{\textbf{F1}} & \multicolumn{1}{c}{\textbf{Acc}} \\
\midrule
\textbf{DS}               & 0.78 & 0.50 & 0.61 & 0.70 & 0.69 & 0.54 & 0.61 & 0.67 \\
\textbf{QC}               & 0.81 & 0.46 & 0.59 & 0.70 & 0.81 & 0.46 & 0.59 & 0.69 \\
\textbf{DS + QC} & 0.82 & 0.50 & 0.62 & 0.72 & 0.75 & 0.57 & 0.65 & 0.71
\\
\bottomrule
\end{tabular}
\end{threeparttable}
}
\begin{tablenotes} 
\small
\item DS represents DeepSeek-V3; QC represents Qwen3-Coder.
\end{tablenotes}
\vspace{-2em}
\end{table}

\ins{
\smallskip
\noindent
\textbf{Generalizability.} 
To assess generalizability across different LLMs, we evaluated \tech{} with a different base LLM, Qwen3-Coder (a recent state-of-the-art model)~\cite{qwen3coder}, and observed comparable results across benchmarks, confirming \tech{}'s generalizability to LLM choice. Table~\ref{tab:different_llm} shows the results. Specifically, precision improved slightly while recall declined, likely due to Qwen3-Coder's stronger reasoning, which benefits reflection but less so for constraint analysis. To validate this, we replaced only the reflection agent with Qwen3-Coder, keeping the rest unchanged. This improved performance (F1-score from 0.61 to 0.62/0.65), reinforces that better reasoning models can enhance the reflection phase. This also highlights \tech{}'s flexibility in integrating different LLMs for specialized roles.
}

\ins{
Beyond LLM choice, \tech{} also generalizes across languages and bug types. Its high-level design is language- and error-agnostic, enabling adaptation to other languages and bug types. It follows a constraint-driven framework: given an error type (e.g., type inconsistency), \tech{} extracts relevant constraints, performs backward analysis along a representative call chain, and generates tests likely to violate those constraints.
The key to generalization lies in defining constraint schemas for target errors, which is often straightforward. For example, to detect Java null pointer dereferences, one can specify that certain variables must be non-null before use and propagate this constraint backward to identify potentially violating contexts. The rest of the pipeline remains unchanged. Adapting to a new language mainly involves adjusting prompt formatting and using available call chain extraction tools, which is widely available in other ecosystems (e.g., Soot for Java).
Overall, \tech{}'s modular, constraint-centric architecture supports extensibility across languages and error types.
}

\begin{lstlisting}[language=Python, caption={A false-positive produced by RTED}, label={listing:fp_example}]
def test_request_httprepr(self):
    class HttpRequest:
        def __init__(self):
            self.url = 'http://example.com'
            self.method = 123
            self.headers = None
            self.body = b''
    http_request = HttpRequest()
    request_httprepr(http_request)
\end{lstlisting}

\ins{
\smallskip
\noindent
\textbf{False-positives produced by \tech{}.} 
Most false positives in \tech{} stem from LLM hallucinations during constraint analysis.
Listing~\ref{listing:fp_example} shows an example.
The LLM generated a mock class \texttt{HttpRequest} with \texttt{self.method = 123}. Passing this object to \texttt{request\_httprepr()} in Scrapy caused a type error, as the function expects request.method to be a string or bytes. The LLM mistakenly inferred method as an integer since it likely conflated method attributes with nearby numeric HTTP status codes/enumerations. To mitigate such issues, we plan to enhance constraint analysis with lightweight type inference/validation.
}

\section{Threats to Validity}
The \textbf{external threat} primarily stems from the generalizability of \tool{}. To mitigate this, we evaluated it on two established benchmarks, BugsInPy and TypeBugs, which span diverse real-world projects ranging from 3k to 316k lines of code. We also applied \tool{} to six recent, large-scale open-source projects, uncovering \found{} previously unknown bugs. For comparison, we selected state-of-the-art techniques in type error detection and LLM-based test generation. The consistent performance improvements across both benchmarks and unseen projects mitigate this threat to some extent.

The \textbf{internal threats} primarily stem from potential implementation errors in \tool{} or the baselines. 
To mitigate this, we used the official implementation of \pyinder{}~\cite{pyinder_artifact} and adapted \tester{} (originally designed for Java) to Python based on its publicly available implementation. 
For \sym{} and \hits{}, which do not have publicly available code, we reimplemented them according to their original papers and validated their correctness with representative examples. \tool{} itself underwent rigorous internal testing and review by two authors.

The \textbf{construct threats} include LLM randomness, data leakage, and metric selection. To control for randomness, we set the LLM temperature to zero in line with existing work~\cite{ase_study,fuzz4all,clast, dan}. To address data leakage, following existing work~\cite{auger}, we checked whether any error-triggering tests generated by \tool{} matched the reference unit tests in the benchmarks. We found that none of the generated tests aligned with the fixed-version tests. Moreover, all baselines use the same underlying LLM as \tool{}, so the performance improvements are not due to the data leakage of LLMs. Additionally,  \tool{} successfully detected \found{} bugs in recently updated open-source projects, which are not included in the LLM’s training data, further mitigating data leakage concerns.
Regarding metrics, we used widely-used metrics, Precision, Recall, F1-score, and Accuracy, to measure effectiveness. 
\ins{
We also evaluated the efficiency of \tool{}. The pipeline starts with Jarvis, which constructs call chains in an average of 14.16s. This is followed by type constraint analysis, test generation, and reflection, adding 77.58s per focal method on average, for a total runtime of 91.74s. Compared to baselines, \tool{} achieves competitive performance: CHATTESTER requires 72.72s, SymPrompt 103.81s, and HITS 153.84s. While these approaches incur similar overheads, they generate substantially more false positives and detect fewer true bugs than \tech{}. Considering the high manual cost of inspecting false positives, \tech{}’s modest runtime overhead is well justified by its higher precision. Moreover, the process can be further accelerated through parallel execution and optimized LLM inference engines such as vLLM~\cite{vllm}.
}

\section{Related Work}

\smallskip
\noindent
\textbf{LLM-based Unit Test Generation.}
LLM-based test generation approaches can be broadly categorized into two types: training-based and prompting-based~\cite{a3test, TELPA, chen2017learning}. Training-based methods, such as ATHENATEST~\cite{athenatest} and A3Test~\cite{a3test}, train LLMs on large-scale datasets of unit tests. While these methods have shown strong performance, they require significant computational resources and large amounts of labeled data. In contrast, prompting-based approaches like \tester{}~\cite{chattester}, \sym{}~\cite{symprompt}, \hits{}~\cite{hits}, and TELPA~\cite{TELPA} guide LLMs using contextual prompts, offering more flexibility and reducing reliance on model fine-tuning.

However, these methods primarily focus on improving test coverage and do not explicitly target bug detection. A recent work by Xin et al. introduces an attention-based mechanism to identify defective methods and guide LLMs to generate bug-revealing tests for Java~\cite{auger}. However, this approach requires a large number of defective method examples with precise error annotations (e.g., faulty lines) for model training, making it costly in terms of time and computational resources. Moreover, such annotated data is difficult to collect in domains like Python type errors.
\textit{In contrast to prior approaches that either focus on coverage improvement or rely on supervised training for general bug detection in Java, our method employs type constraint analysis to guide LLMs in generating tests that are more likely to trigger Python type-related bugs, and incorporates a reflection phase to mitigate false positives.}

\smallskip
\noindent
\textbf{Python Type Analysis.}
Several static type analysis tools for Python have been proposed, including Pyre~\cite{pyre}, with support for gradual typing and custom annotations; Pyright~\cite{pyright}, a fast and feature-rich type checker; Mypy~\cite{mypy}, one of the earliest static type checkers in the Python community; and Pytype~\cite{pytype}, which does not require explicit type annotations.
Recently, \pyinder{}, which is discussed and compared in our evaluation, builds upon the existing tools to improve type error detection and achieves state-of-the-art performance among static analyzers.
\textit{Different from these work, \tool{} takes a dynamic testing approach via step-by-step type constraint analysis and reflection mechanism to iteratively guide the test generation process, enabling more precise detection of type errors.
}

\smallskip
\noindent
\textbf{Python Type Inference.}
Static analysis has been extensively applied to infer types in Python programs using techniques such as constraint-based inference~\cite{infer1} and abstract interpretation~\cite{infer2, infer3}. More recently, LLM-based methods have emerged for Python type inference. For example, TypeGen~\cite{typegen} combines lightweight static analysis with in-context learning, crafting few-shot Chain-of-Thought (CoT) prompts to enhance type inference performance. TIGER~\cite{tiger} adopts a two-stage generate-then-rank framework to handle Python’s complex and diverse type system more effectively.
\textit{Unlike these approaches, which aim to infer precise concrete types, typically at the method level. Our approach analyzes type constraints at the call-chain level and represent type constraints in a unified form instead of concrete types.}

\section{Conclusion}
In this paper, we present \tool{}, a novel type-aware unit test generation framework for effectively detecting type errors. \tool{} combines step-by-step type constraint analysis and reflective validation to guide test generation while minimizing false positives. Our evaluation on two widely-used benchmarks (i.e., BugsInPy and TypeBugs) shows that \tool{} can detect 22$\sim$29 more benchmarked type errors than state-of-the-art techniques, including \pyinder{}, \tester{}, \sym{}, and \hits{}.
\tool{} is also capable of producing fewer false positives, achieving an improvement of 173.9\%$\sim$245.9\% in precision.
Furthermore, \tool{} detects \found{} previously unknown type errors in large-scale real-world Python projects, demonstrating its effectiveness and generalizability in practical application.

\section*{Acknowledgment}
We thank all the ASE anonymous reviewers for their valuable comments.
This work was supported by the National Key Research and Development Program of China (Grant No. 2024YFB4506300), the National Natural Science Foundation of China (Grant Nos. 62322208, 62232001), and the Emerging Frontiers Cultivation Program of Tianjin University Interdisciplinary Center.

\balance
\bibliographystyle{IEEEtran}
\bibliography{ref}

\end{document}